\documentclass[twocolumn]{svjour3}          
\smartqed  
\usepackage{graphicx}
\usepackage{mathptmx}      
\usepackage{latexsym}
\usepackage{amsmath}
\usepackage{amssymb}
\usepackage{soul}
\usepackage[colorlinks=true,citecolor=blue,linkcolor=red,linktocpage=true]{hyperref}
\journalname{Journal of Computational Electronics}
\begin{document}

\title{
Floquet scattering matrix approach to the phase noise of a single-electron source in the adiabatic regime 
}


\author{
Michael  Moskalets 
}


\institute{
M. Moskalets \at
Department of Metal and Semiconductor Physics, NTU ``Kharkiv Polytechnic Institute", 61002 Kharkiv, Ukraine \\
              Tel.: +38-057-7076601\\
              Fax: +38-057-7001564\\
              \email{michael.moskalets@gmail.com}           
}

\date{Received: date / Accepted: date}

\maketitle

\begin{abstract}
We give the basic elements of the Floquet scattering matrix approach \cite{Moskalets:2011cw} to the dynamic quantum transport in mesoscopic and nanoscopic conductors. 
We use the scattering formalism to discuss the noise power spectrum of a single electron source working in the adiabatic regime and emitting particles into a chiral electron  waveguide.  
The noise power is found to be quadratic at low frequencies and exponentially suppressed at high frequencies. 
\end{abstract}

\section{Introduction}
\label{intro}

Recent progress achieved with high-speed single-electron sources   \cite{Blumenthal07,Feve:2007jx,Kaestner:2008gv,Fujiwara:2008gt,Dubois:2013ul} have opened fascinating perspectives for the development of quantum electronics. 
On one side, it advances essentially the field of electrical quantum metrology \cite{Giblin:2012cl,Hohls:2012gf,Pekola:2012ti,Jehl:2013wd} by offering a real possibility to close the quantum metrological triangle \cite{Likharev:1985ul,Flowers:2004ei,Feltin:2009fo,Scherer:2012fe}. 
On the other side, the on-demand injection of a single-particle state makes it possible to develop an electronic analogue \cite{Grenier:2011js} of quantum optics having a great potential for quantum information processing \cite{Bennett:2000kl}. 
For example, the tunable two-particle exchange was already proposed \cite{Olkhovskaya:2008en}  and measured \cite{Bocquillon:2013dp}. 

The difference of a single electron source (SES) from a biased metallic contact, which is more conventional for electronics, is that SES emits electrons well separated in space and time. 
It allows to manipulate in a controllable manner single particles and to engineer few-particle states necessary for storing and processing information.   
With this respect the crucial question is whether the emitted state is a genuine single-particle state or not. 
To answer this question the high-frequency noise of currents associated with a single-electron emission was measured \cite{Mahe:2010cp}.
It was demonstrated that there exists a fundamental lower limit to the noise due to a quantum uncertainty in the time of emission. \cite{Mahe:2010cp,Albert:2010co} 
The source, regularly emitting single particles, nevertheless generates fluctuating currents. 
This limiting noise is referred to as the phase noise. 
If, in addition, the emission is irregular or involves more than a single particle per time, the noise gets enhanced over this limit. 
The numerical calculations based on the Floquet scattering theory show an excellent agreement with experimental data.  \cite{Parmentier:2012ed}

In this contribution our aim is to develop an analytical theory of the phase noise of a single-electron emitter working in the adiabatic regime. 
For this purpose we use the Floquet scattering theory of a time-dependent quantum transport. \cite{Moskalets:2002hu}
In the adiabatic regime the source is driven by an AC voltage, which varies  slowly compared to the internal dynamics of an electron source. 
Therefore, the source stays in a close to equilibrium state while emitting electrons. 
In contrast, in the non-adiabatic regime of Refs.~\cite{Mahe:2010cp} the source is driven by a pulsed potential and, therefore, it emits electrons during a highly non-equilibrium, transient process.  
Despite of this difference we find the noise spectrum to be qualitatively similar: 
It shows a universal quadratic behavior at low frequencies that reflects a probabilistic tunneling dynamics of an electron leaving the source. 
At high frequencies the phase noise is suppressed.  
Using analytical calculations we show that in the adiabatic emission regime this suppression admits both a time-based and an energy-based interpretation. 
From the former point of view, the phase noise gets suppressed as a result of a current spectrum suppression at high frequencies. 
From the latter point of view, the phase noise suppression is due to decrease in the number of electrons able to emit high-frequency energy quantum.

The paper is organized as follows. 
The basic elements of the Floquet scattering matrix formalism are presented in Sec.~\ref{scat}. 
In Sec.~\ref{encahb} the scattering amplitude of a single-particle emitter, the current generated, and the current correlation function are derived.  
In Sec.~\ref{adia} the adiabatic working regime is discussed. 
The excess noise spectrum is calculated and analyzed in Sec.~\ref{ffn}. 
We conclude in Sec.~\ref{concl}.

\section{Floquet scattering matrix formalism} 
\label{scat}

\begin{figure}[t]
\begin{center}
\includegraphics[width=80mm]{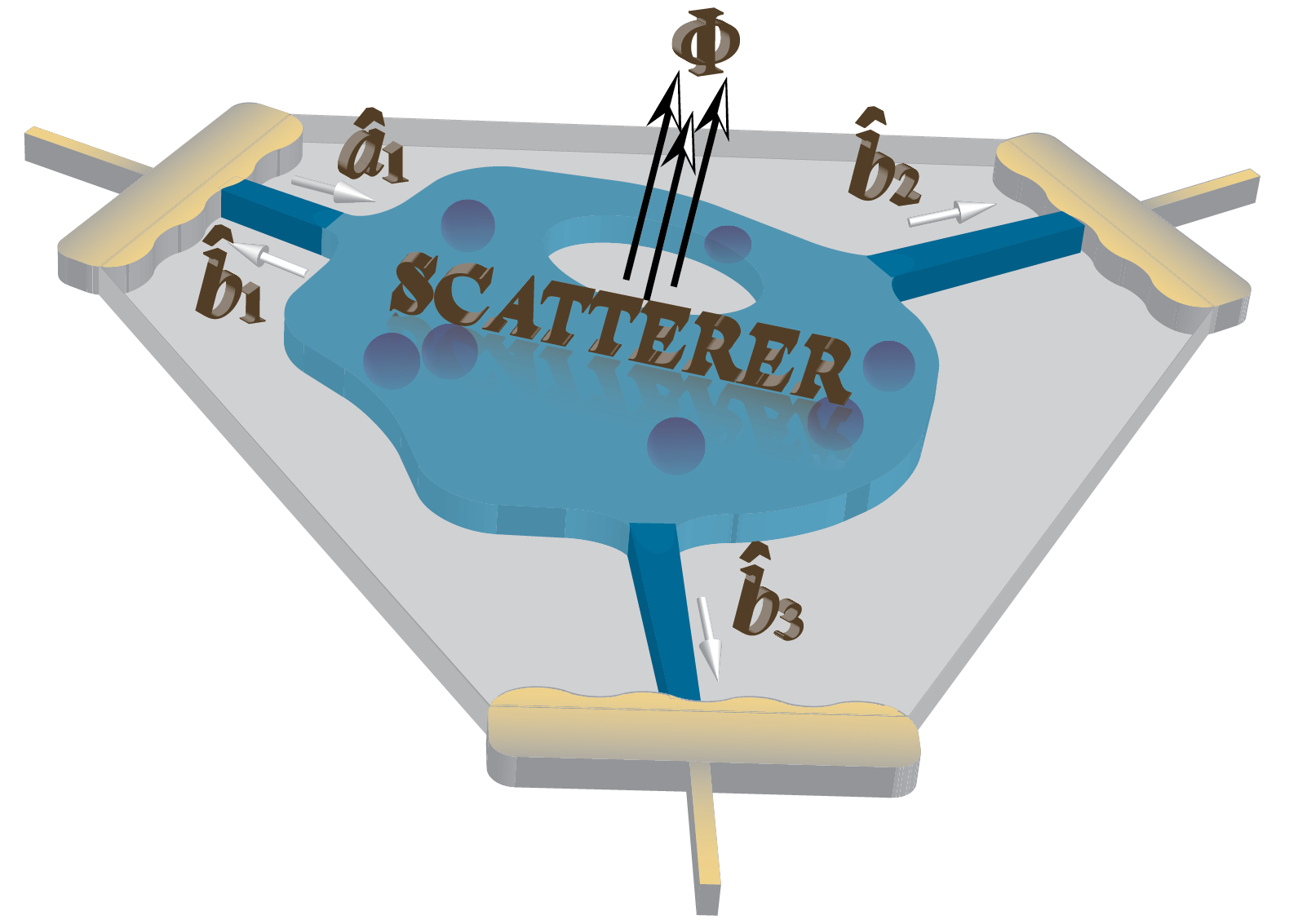}
\caption{(Color online)
Mesoscopic scatterer  connected to three contacts, $M_{r} = 3$. An electron originated from the first contact (operator $\hat a_{1}$) is shown to be transmitted to the second and the third contacts (operators $\hat b_{2}$ and $\hat b_{3}$, respectively) or reflected back to the first contact (operator $\hat b_{1}$). The scatterer shown includes impurities (dark circles) and is threaded by the magnetic flux $\Phi$. 
}
\label{fig1}
\end{center}
\end{figure}

Here we closely follow the notations introduced in Ref.~\cite{Moskalets:2011cw}.  
Let us consider a mesoscopic sample (a scatterer), connected to several, $M_{r}$,  metallic contacts via one-dimensional leads, Fig.~\ref{fig1}.  
The metallic contacts play the role of electron reservoirs \cite{Landauer:1988vs}, which we label with greek letters, $\alpha = 1,\dots, M_{r}$. 
We consider a sample as mesoscopic if electrons propagating through it preserve phase coherence. 
Electrons in each reservoir $\alpha$ are assumed to be in equilibrium.
An electron from any reservoir $\alpha$ can come to the sample. 
Then it will be either transmitted to another reservoir $\beta$ or reflected back to the same reservoir.  
If the reservoirs have different chemical potentials, $\mu_{\alpha} \ne \mu_{\beta}$, and/or temperatures, $T_{\alpha} \ne T_{\beta}$, then electrons scattered between different reservoirs become non--equilibrium at the destination. 
This results in appearance of currents through the sample driven by the chemical potential (and/or temperature) difference.  
Another possibility to drive the system out of equilibrium is to act directly onto the sample and to change its properties in time.
\cite{Switkes:1999fj} 
For instance, one can vary a magnetic flux penetrating a scatterer, see Fig.~\ref{fig1}, or an electrostatic potential induced by the nearby gate.  
After scattering off a dynamic sample, an electron changes its energy and becomes non--equilibrium even if it is scattered back to the reservoir it came from.

\subsection{Current operator and current correlation function}
\label{coccf}

Here we outline shorty how to calculate a current flowing through the system. 
We use a charge current as an example. 
However the same method is appropriate to address any kind of a flux, energy, spin, etc.      
First, we introduce second quantization operators $\hat a_{\alpha}^{\dag}(E),\,\hat a_{\alpha}(E)$ which create, annihilate an electron with energy $E$ incoming to the scatterer from the reservoir $\alpha$.  
As we already mentioned, the incoming electrons are in equilibrium and, therefore, the quantum-statistical average of the product of creation and annihilation operators is given by the Fermi distribution function $f_{\alpha}(E)$ of the corresponding reservoir,
\begin{equation}
\left\langle \hat a_{\alpha}^{\dag}(E) a_{\alpha}(E^{\prime})  \right\rangle = f_{\alpha}(E) \delta(E-E^{\prime})\,.
\label{01}
\end{equation}
Here $\delta(E - E^{\prime})$ is the Dirac delta function and the Fermi distribution function is, 
\begin{eqnarray}
f_{\alpha}(E) = \frac{ 1}{ 1 + e^{ \frac{E - \mu_{\alpha} }{k_{B} T_{\alpha} } } } \,,
\label{fermi}
\end{eqnarray}
where $k_{B}$ is the Boltzmann constant, $\mu_{\alpha}$ is the chemical potential, and $T_{\alpha}$ is the temperature of the reservoir $\alpha$. 
We stress the reservoirs are assumed to be macroscopic and their states are not changed by the coupling to a small mesoscopic system. 

Second, we introduce operators $\hat b_{\alpha}^{\dag}(E),\,\hat b_{\alpha}(E)$ for electrons scattered off the sample into the lead $\alpha$. 
Since an electron scattered to the lead  $\alpha$ can come from any other lead, the operator $b_{\alpha}$ can be related to all the operators  $a_{\beta}$, $\beta = 1, \dots, M_{r}$. 
The coefficients in such a relation are the elements $S_{\alpha\beta}$ of the scattering matrix ${\bf S}$. 
If the scatterer is stationary the scattering matrix depends on only one energy argument, ${\bf S}(E)$. 
In contrast the scattering matrix for a dynamic sample depends on two energy arguments, ${\bf S}(E_{2},E_{1})$. 
Since the energy of incoming, $E_{1}$, and scattered, $E_{2}$, electron can be different. 

Here we are interested in an important case when the scatterer is driven periodically in time with frequency $\Omega$. 
\cite{Platero:2004ep,Arrachea:2007ew}
In this case the energy, which an electron can pick up during scattering, is quantized in units of $\hbar\Omega$. 
We call the corresponding scattering matrix as the {\it Floquet} scattering matrix, ${\bf S}_{F}$.
Then the relation between  $b-$ and $a-$operators reads, \cite{Moskalets:2002hu}
\begin{equation}
\hat b_{\alpha}(E) = \sum\limits_{\beta = 1}^{M_{r}} \sum\limits_{n = - \infty}^{\infty} S_{F,\alpha\beta}\left( E, E_{n} \right) \hat a_{\beta}(E_{n})\,.
\label{02}
\end{equation}
Here $E_{n} = E + n\hbar\Omega$. 
Strictly speaking the lower bound in the sum over $n$ is restricted by the requirement, $E_{n} > 0$, in order that incoming electrons are propagating. 
However if the energy quantum $\hbar\Omega$ is much smaller than the bandwidth, the Fermi energy, the lower bound can be put to minus infinity, what we assume  in equation (\ref{02}). 
The current conservation requires the Floquet scattering matrix to be unitary. \cite{Moskalets:2011cw,Moskalets:2002hu}
For a single orbital channel case of interest here, the unitarity condition reads, \cite{Moskalets:2004bo}
\begin{eqnarray}
\sum\limits_{p = -\infty}^{\infty} S_{F}^{*}(E_{p}, E_{m}) S_{F}(E_{p}, E_{n}) = \delta_{m,n} \,,
\nonumber \\
\label{uni} \\
\sum\limits_{p = -\infty}^{\infty} S_{F}^{*}(E_{m}, E_{p}) S_{F}(E_{n}, E_{p}) = \delta_{m,n} \,,
\nonumber 
\end{eqnarray}
where $\delta_{m,n}$ is the Kronecker symbol. 

If the relevant energy scales (such as the applied voltage, the temperature difference, the energy quantum $\hbar\Omega$, etc.) characterizing the non-equilibrium state are all small compared to the Fermi energy $\mu$, then the current operator in lead $\alpha$ reads, \cite{Buttiker:1992vr}
\begin{equation}
\hat I_{\alpha}(t) = \frac{{\rm e} }{h}\! \iint\!\! dE dE^{\prime} e^{i\frac{E-E^{\prime}}{\hbar}t} \!\left\{ \hat b_{\alpha}^{\dagger}(E) \hat b_{\alpha}(E^{\prime}) - \hat a_{\alpha}^{\dagger}(E) \hat a_{\alpha}(E^{\prime})   \right\} ,
\label{Eq16}
\end{equation}
where ${\rm e} < 0$ is the elementary charge and $h$ is the Planck constant. 
The quantum statistical average of this operator over the equilibrium state of electrons in the reservoirs gives a time-dependent current, $I_{\alpha}(t) = \left\langle \hat I_{\alpha}(t)  \right\rangle$, flowing into the lead $\alpha$. 
To perform such averaging  we use Eqs.~(\ref{02}) and (\ref{01}). 

Here we are interested in fluctuations of this current, $\Delta \hat I_{\alpha} = \hat I_{\alpha} - \left\langle \hat I_{\alpha } \right\rangle$.  
We characterize such fluctuations with the help of a current correlations function. In the frequency domain it reads as follows, \cite{Blanter:2000wi}
\begin{equation}
{\rm P}_{\alpha\beta}(\omega_{1},\omega_{2}) = \frac{1}{2} \left\langle \Delta \hat I_{\alpha}(\omega_1) \Delta \hat I_{\beta}(\omega_2) + \Delta \hat I_{\beta}(\omega_2)  \Delta \hat I_{\alpha}(\omega_1) \right\rangle ,
\label{Eq2_3_09}
\end{equation}
where 
\begin{equation}
\hat I_{\alpha}(\omega) = {\rm e} \int\limits_{0}^{\infty} dE \left\{ \hat b_{\alpha}^{\dagger}(E) \hat b_{\alpha}(E+\hbar\omega) - \hat a_{\alpha}^{\dagger}(E) a_{\alpha}(E + \hbar\omega) \right\} ,
\label{Eq2_3_10}
\end{equation}
is the Fourier transform of a current operator $\hat I_{\alpha}(t)$, Eq.~(\ref{Eq16}).

\section{Single electron source}   
\label{encahb}

\begin{figure}[t]
\begin{center}
\includegraphics[width=80mm]{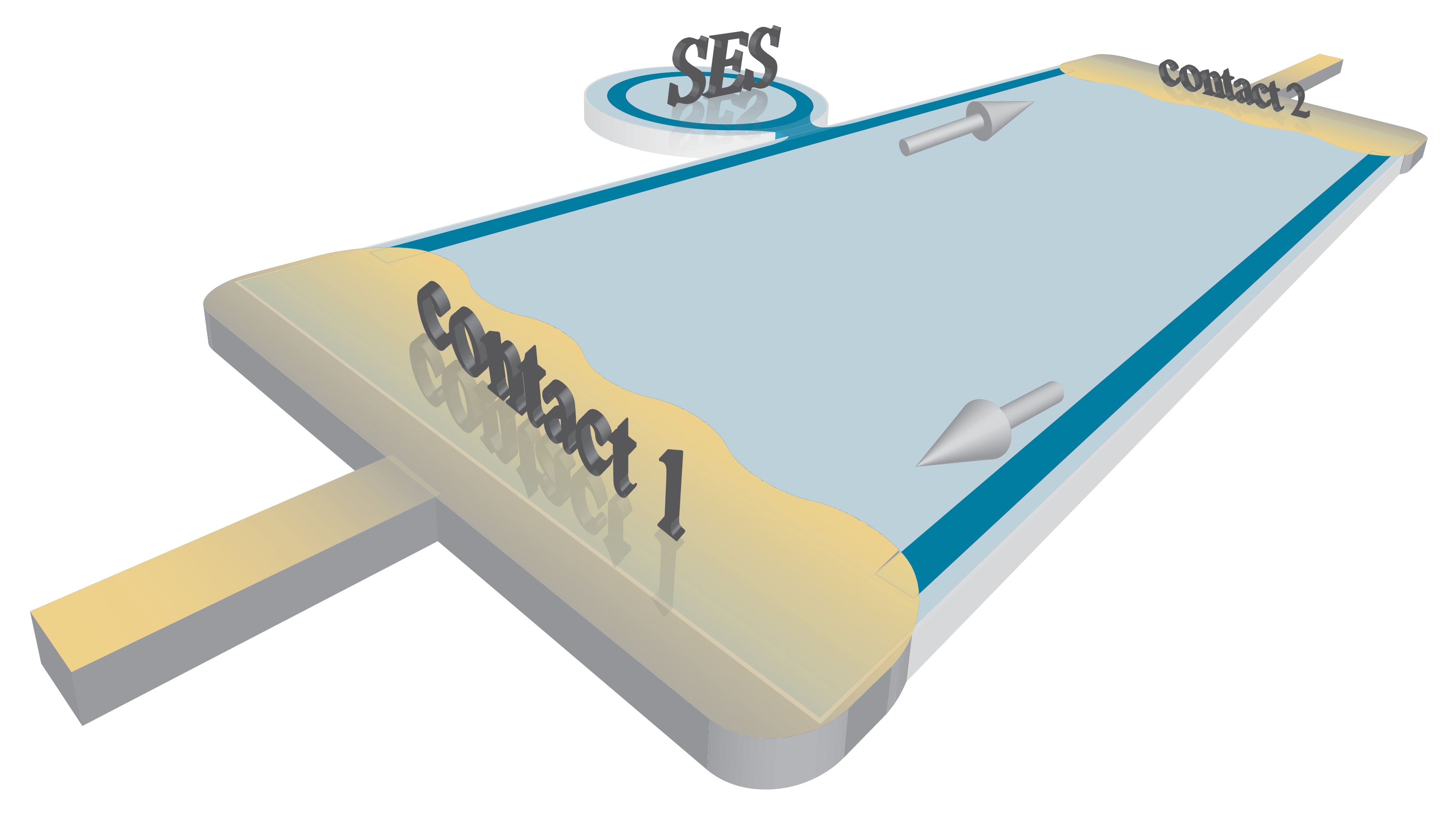}
\caption{(Color online) Single-electron source (SES). 
The mesoscopic capacitor, a circular edge state, is attached to the Hall bar. The arrows indicate the direction of movement of electrons in the edge states (the blue strips). In the capacitor electrons move clockwise. The capacitor is driven by the periodic potential applied to the nearby metallic gate (not shown). The electrons (holes) emitted by the capacitor flow to the metallic contact $2$ where the current $I_{2}$ is measured.  
}
\label{fig2}
\end{center}
\end{figure}

Below, to be specific, we concentrate on a single-electron source used in Ref.~\cite{Mahe:2010cp}. 
The cartoon illustrating the features of the experimental set-up essential for us is shown in Fig.~\ref{fig2}.
The source is made of a mesoscopic  capacitor \cite{Buttiker:1993wh} in a two-dimensional electron gas in the quantum Hall effect regime \cite{Klitzing:1980wf}. 
In this regime electrons propagate along chiral states located close to the edges of a sample
\cite{Halperin:1982tb,Buttiker:1988vk} 
The capacitor, a small cavity with a circular edge state, is side-attached to the linear edge state playing the role of an electron waveguide. 
The length of a circular state is small such that the electron spectrum is quantized. 
The periodic potential of the top gate surrounding the capacitor (not shown in Fig.~\ref{fig2}) drives its energy levels up and down. 
When some quantum level of the cavity rises above the Fermi level in the waveguide, an electron is emitted. 
In the opposite process, when the level in the cavity sinks below the Fermi energy, an electron enters the cavity, a hole is emitted into the waveguide. 
Since after completion of a period no net charge is emitted, such a cavity is called a capacitor.

The current and its fluctuations at the contact $2$ are of interest here. 
To calculate them we need the $b_{2}(E)$ operator, which is (see, Eq.~(\ref{02})),
\begin{equation}
\hat b_{2}(E) =  \sum\limits_{n = - \infty}^{\infty} S_{F,21}\left( E, E_{n} \right) \hat a_{1}(E_{n})\,.
\label{02add}
\end{equation}
The scattering element $S_{F,22} = 0$ since an electron motion is unidirectional. 
The non-zero scattering element is 
\begin{eqnarray}
S_{F,21}\left( E, E_{n} \right) = e^{ik(E) L_{2c}} S_{F}\left( E, E_{n} \right) e^{ik(E_{n}) L_{c1}} . 
\label{add01}
\end{eqnarray}
Here $L_{2c}$ ($L_{c1}$) is a distance between the cavity and the contact $2$ ($1$); $k(E)$ is a wave number of an electron with energy $E$ in the waveguide;   
$S_{F}(E,E_{n})$ is the Floquet scattering matrix of the cavity. 
The exponential factors in Eq.~(\ref{add01}) describe a free propagation of electrons from the contact $1$ to the cavity and from the cavity to the contact $2$. 
They are irrelevant for subsequent calculations. 
In contrast, the scattering off the cavity is crucial. 
The element $S_{F}(E,E_{n})$ describes a forward scattering of electrons in the chiral waveguide the cavity is coupled to. 
During such a scattering an electron energy is changed from $E_{n} = E + n\hbar\Omega$ to $E$ due to interaction with a dynamic cavity. 
Now we show how the Floquet scattering matrix of the cavity is calculated. 
We stress it is a matrix in the energy space.  

\subsection{Scattering amplitude of the SES}
\label{ses}

The cavity is modeled \cite{Pretre:1996uw,Gabelli:2006eg} as a single-channel chiral state of the length $L$ with a uniform time-periodic potential $U(t) = U(t+{\cal T})$.  
This state is coupled with the help of a quantum point contact (QPC) to a single-channel linear edge state playing the role of an electron waveguide. 
The QPC is characterized by the energy independent reflection and transmission coefficients.  
The problem of scattering of electrons in the linear edge state off this scatterer can be solved exactly and the Floquet scattering matrix of the cavity can be calculated. \cite{Moskalets:2008fz} 

To simplify calculations, it is convenient to represent the Floquet scattering matrix elements as the Fourier transform of some auxiliary scattering amplitude dependent on one energy and one time arguments. 
Depending on which energy, for incoming or outgoing electrons, is kept fixed, this amplitude is referred to as $S_{in}(t,E)$ or $S_{out}(E,t)$, respectively. \cite{Moskalets:2008ii}
For the purposes of this paper it is convenient to work with the latter amplitude, which is defined as follows,
\begin{equation}
S_{F}\left(E, E_{n}\right) = S_{out, -n}(E)  \equiv \int\limits_{0}^{\cal T} \frac{dt}{\cal T} e^{-in\Omega t} S_{out}(E,t) \,.
\label{03}
\end{equation}
For the model of interest here the amplitude $S_{out}(E,t)$  has the following representation in terms of particular amplitudes $S^{(q)}(t)$~: \cite{Moskalets:2008fz}
\begin{eqnarray}
S_{out}(E,t) &=& \sum\limits_{q=0}^{\infty}  S^{(q)}(t)\,, 
\nonumber \\
\label{04} \\
S^{(0)} &=& r\,, 
\nonumber \\
S^{(q>0)}(t) &=& \bar t^2\, r^{q-1}\,e^{iqkL}\, e^{-i\Phi_{q}(t)}\,, 
\nonumber \\
\Phi_q(t) &=& \frac{{\rm e} }{\hbar}\int\limits_{t}^{t+q\tau} dt^\prime U(t^\prime) \,. 
\nonumber 
\end{eqnarray}
Here $k\equiv k(E)$ is a wave number of an electron with energy $E$ in the cavity; $r$ and ${\bar t}$ are reflection and transmission coefficients of the QPC connecting the cavity and an electron waveguide. 
The partial amplitude $S^{(q)}(t)$  is an amplitude for an electron to enter the cavity at time $t$ and to leave the cavity after completing $q$ full turns (of duration $\tau$ each). 
The energy of an electron leaving the cavity is $E$.
The quantity $\Phi_q(t)$ is a time-dependent phase acquired by an electron during such a propagation. 
The potential energy ${\rm e}U$ is assumed to be small compared to an electron energy  $E$ that allows to separate the time-dependent phase $\Phi_{q}(t)$ and the orbital phase $qkL$. 
This general solution is not restricted to any particular amplitude and/or time--dependence of the driving potential $U(t)$.

\subsection{Current}   
\label{it}

We substitute Eq.~(\ref{02add}) into Eq.~(\ref{Eq16}) and calculate a time dependent current, $I_{2}(t) = \left\langle \hat I_{2}(t)  \right\rangle$, measured at the contact $2$  (see Fig.~\ref{fig2}), 
\begin{eqnarray}
I_{2}(t) &=& \sum\limits_{\ell=-\infty}^{\infty} e^{- i \ell \Omega t} \frac{{\rm e} }{h } \int\limits_{0}^{\infty} dE 
\label{it01} \\
&& \times \sum\limits_{n =-\infty}^{\infty} S_{F}^{*}\left(E_{n}, E  \right) S_{F}\left(E_{n+\ell}, E  \right) \left\{ f(E) - f\left( E_{n} \right) \right\} . 
\nonumber
\end{eqnarray}
Here we suppose that both contacts, $\alpha=1,2$, have the same chemical potential, $\mu_{1} = \mu_{2} \equiv \mu$, the same temperature, $T_{1}= T_{2} \equiv T$, and  thus are characterized by the same Fermi distribution function, $f_{1}(E) = f_{2}(E) \equiv f(E)$.

\subsection{Current correlation function}   
\label{ccf}

In the set-up shown in Fig.~\ref{fig2} the current auto-correlation function is measured at the contact $2$. 
We use Eq.~(\ref{Eq2_3_09}) and find, \cite{Moskalets:2007dl} 
\begin{subequations}
\label{001}
\begin{eqnarray}
{\rm P}_{22}(\omega_{1},\omega_{2}) = \sum\limits_{\ell = -\infty}^{\infty} 2\pi \delta\left( \omega_{1} + \omega_{2} - \ell \Omega \right) {\cal P}_{\ell}(\omega) \,,
\label{001A}
\end{eqnarray}
where
\begin{eqnarray}
{\cal P}_{\ell }(\omega_{1},\omega_{2})  = \frac{{\rm e}^{2}  }{h} \int\limits_0^\infty  dE\,\bigg\{ \delta _{l0}\, F \left( {E,E + \hbar \omega_{1} } \right) \quad \quad\quad \quad
\label{001B} 
\\
 + \sum\limits_{n,m,p =  - \infty }^\infty   F \left( {E_{ \ell + n} ,E_m  + \hbar \omega_{1} } \right)S_{F}^{} \left( {E_{ \ell + p} ,E_{ \ell + n} } \right)S_{F}^* \left( E,E_{ \ell + n} \right) \nonumber \\
\nonumber \\
\times  S_{F}^{} \left( {E + \hbar \omega_{1},E_m  + \hbar \omega_{1} } \right)S_{F }^* \left( {E_p  + \hbar \omega_{1},E_m  + \hbar \omega_{1} } \right)\!\! \bigg\} , 
\nonumber
\end{eqnarray}
\end{subequations}
and 
\begin{equation}
F(E_{1},E_{2}) = \frac{ f(E_{1})\left [ 1 - f(E_{2}) \right] + f(E_{2})\left [ 1 - f(E_{1}) \right] }{2} \,.
\label{002}
\end{equation}
The current correlation function is non-zero even in equilibrium, when the driving potential is switched off. 
This is due to the quantum noise. \cite{Gardiner:2000wq,Clerk:2010dh} 
The important feature of the present set-up is the possibility to completely subtract the  equilibrium quantum noise and to extract the contribution due solely to the emitted particles.   
For this purpose the excess noise was measured. \cite{Mahe:2010cp}
The excess refers to the difference between the current correlation functions  measured with the source on and off. 
We will indicate the excess noise with the superscript $(ex)$. 

Here we discuss only the  ${\ell} = 0$ component of the excess auto-correlation function, which was measured in Ref.~\cite{Mahe:2010cp}: 
\begin{eqnarray}
{\cal P}_{0}^{(ex)}(\omega) = \frac{{\rm e}^{2}  }{h} \int\limits_0^\infty  dE  \sum\limits_{m =  - \infty }^\infty    F ( {E ,E_{m}  + \hbar \omega } )  
\label{33} \\ 
 \times \left\{ \left | \Pi_{m}(E,\omega) \right |^{2}  - \delta_{m,0}  \right\} ,
 \nonumber
\end{eqnarray} 
where 
\begin{eqnarray}
\Pi_{m}(E,\omega) &=& \sum_{q=-\infty}^{\infty} S_{F} \left( E_{q},E \right) S_{F}^* \left( E_{q}+ \hbar \omega ,E_{m} + \hbar \omega \right) .
\label{34}
\end{eqnarray}
An equation similar to Eq.~(\ref{33}) was found in Ref.~\cite{Parmentier:2012ed}.

\section{Adiabatic emission}
\label{adia}

We suppose that the potential $U(t)$ changes slowly compared to the average time an electron spends in the cavity, the dwell time, 
\begin{eqnarray}
\tau_{D} = \frac{\tau }{ \rm T} \,.
\label{dwt}
\end{eqnarray}
Here ${\rm T} = |{\bar t}|^{2}$ is the transmission probability of the QPC connecting the  cavity and an electron waveguide; $\tau$ is a time of a single turn around the cavity.  In such a case we can keep $U(t)$ constant while calculating the time-dependent phase,
\begin{eqnarray}
\Phi_q(t) = \frac{\rm e }{\hbar } q\tau U(t)\,.
\label{tdph}
\end{eqnarray} 
Then we can easily sum up over $q$ in Eq.~(\ref{04}) and find $S_{out} (E,t) = S(E,U(t))$, where  
\begin{equation}
S(E,U(t)) = - e^{ i(\phi(E,t) + \theta_{r}) } \frac{ 1 -   \sqrt{R} e^{-i\phi(E,t)} }{ 1 - \sqrt{R} e^{i\phi(E,t)} } \,.
\label{06}
\end{equation}
Here we write the reflection coefficient as $r = \sqrt{R}\exp(i\theta_{r})$ and introduce the phase $\phi(E,t)$ acquired by an electron during one turn,
\begin{eqnarray}
 \phi(E,t) &=& \theta_{r} + k_{\mu}L + 2\pi \frac{ E  - \mu - {\rm e}U(t)  }{ \Delta } \,, 
 \label{05} 
\end{eqnarray}
where $\Delta = h/\tau$ is the level spacing in the cavity and $k_{\mu} = k(\mu)$ is the  wave number of an electron with the Fermi energy $\mu$. 
In above equation we expand the orbital phase, $kL = k_{\mu}L + 2\pi \left( E-\mu \right)/\Delta$ close to the Fermi energy $\mu$. 
This expansion is  exact for a linear electron dispersion, $E(k)$. 
For a nonlinear dispersion it is a good approximation if the relevant energies are close to the Fermi energy, $|E-\mu| \ll \mu$.  
Note within this approximation the time of one turn around the cavity $\tau$ is energy independent. 
Respectively, the level spacing $\Delta$ is also energy independent, hence the spectrum of electrons in the cavity is equidistant.  

The amplitude $S(E,U)$, Eq.~(\ref{06}), is nothing but the scattering amplitude of a cavity with a fixed potential $U$. 
Since now the potential is time-dependent, the amplitude $S(E,U(t))$ is referred to as the {\it frozen} scattering amplitude, meaning a scattering amplitude with a potential frozen at time $t$. 
This amplitude is manifestly unitary, $|S(E,U(t))|^{2}=1$, as the current conservation requires.

\subsection{Quantized emission regime} 
\label{qer}

If the transmission of the QPC is small, ${\rm T}\ll 1$, the quantum levels in the cavity are well defined, i.e., their width $\delta$ is small compared to the level spacing $\Delta$. 
One can model such levels as the Breit-Wigner resonances.  \cite{Breit:1936ud}
With varying potential $U(t)$ the position of the levels in the cavity are  changed. 
We suppose the only one level in the cavity crosses the Fermi level. 
Close to the time of an electron emission (when the level in the cavity rises above the Fermi level in an electron waveguide) the scattering amplitude can be represented as follows (up to the irrelevant phase factor), \cite{Olkhovskaya:2008en}
\begin{eqnarray}
S(E,t) =  \frac{t - t_{-}(E) + i \Gamma_{\tau}(E) }{t - t_{-}(E) - i \Gamma_{\tau}(E) } \,,
\label{06-1}
\end{eqnarray}
where $t_{-}(E)$ is a time when the middle of an energy level is equal to $E$; the parameter $\Gamma_{\tau}(E)$ is a time interval, during which the level having a width $2\delta$   crosses the energy $E$. 
For an equidistant spectrum with level spacing $\Delta$ the level half-width is $\delta = {\rm T} \Delta/(4\pi)$ and $\Gamma_{\tau} = \delta/ |{\rm e} \partial_{t} U\left( t_{-} \right)| $ with $\partial_{t}$ indicating a time derivative. 
Close to the time of a hole emission, $t_{+}$, the complex conjugate to Eq.~(\ref{06-1}) should be considered.

\subsection{Current in the adiabatic regime}
\label{car}

The adiabatic regime implies that the Floquet scattering matrix changes only a little with energy on the scale put by the frequency, i.e., on the scale $\hbar\Omega$. \cite{Moskalets:2002hu} 
To the leading order in $\Omega$ its matrix elements are the Fourier coefficients of the frozen scattering amplitude, see Eqs.~(\ref{03}) and (\ref{06}). 
Thus we can use $S_{F}^{*}\left(E_{n}, E  \right) \approx S_{n}^{*}(E)$ and $S_{F}\left(E_{n+\ell}, E  \right) \approx S_{n+\ell}(E)$. 
Moreover, expanding the Fermi function difference, $f(E) - f(E_{n}) \approx \left( - \partial f/\partial E \right) n \hbar \Omega$, we rewrite Eq.~(\ref{it01}) as follows, 
\begin{eqnarray}
I_{2}(t) = \frac{- i {\rm e} }{2\pi } \int\limits_{0}^{\infty} dE \left( - \frac{\partial f }{\partial E } \right)\ S(E,t) \frac{\partial S^{*}(E,t) }{\partial t } .
\label{it02}
\end{eqnarray}
At low temperatures, $k_{B} T \leq \hbar\Omega$, we can keep the scattering amplitude as energy independent, $S(E,t) \approx S(\mu,t)$, while integrating over energy in Eq.~(\ref{it02}). 
Using Eq.~(\ref{06-1}) we find a time-dependent current generated by a single-particle emitter in the adiabatic regime: \cite{Olkhovskaya:2008en}
\begin{eqnarray}
I_{2}(t) = \frac{{\rm e} }{\pi } \left\{ \frac{\Gamma_{\tau} }{ \left( t - t_{-} \right)^{2} + \Gamma_{\tau}^{2}  } - \frac{\Gamma_{\tau} }{ \left( t - t_{+} \right)^{2} + \Gamma_{\tau}^{2}  }  \right\} .
\label{it03}
\end{eqnarray}
Here all quantities are calculated at the Fermi energy: $t_{\mp} = t_{\mp}(\mu)$ and $\Gamma_{\tau} = \Gamma_{\tau}(\mu)$. 
In above equation the current $I_{2}(t)$ is given for a single period, $0 < t < {\cal T}$ and should be periodically extended to other times. 
Note for the adiabatic regime to be held the following inequality should be satisfied, \cite{Splettstoesser:2008gc}
\begin{eqnarray}
\Gamma_{\tau} \gg \tau_{D} \,, 
\label{adcond}
\end{eqnarray}
i.e., the current pulse duration should be much larger than the dwell time. 

The current $I_{2}(t)$ consists of two pulses of a Lorentzian shape:  
The first, positive pulse corresponds to an emitted electron (it carries a charge ${\rm e}$), while the second, negative pulse corresponds to an emitted hole (it carries a charge $-{\rm e}$).  
Consequently the current spectrum is bounded. 
The discrete Fourier transform of the current reads,
\begin{eqnarray}
I_{2,n} = \frac{{\rm e} }{\cal T } e^{- |n| \Omega \Gamma_{\tau} } \left\{ e^{i n \Omega t_{-}} + e^{i n \Omega t_{+}} \right\} .  
\label{it04}
\end{eqnarray}
Here we took into account that $\Gamma_{\tau} \ll {\cal T}$. 
As we will see below the exponential decay of a current spectrum results in an  exponential decay of an excess noise spectrum.   

\section{Excess noise in the adiabatic regime}
\label{ffn}

As we will show, the excess noise in the adiabatic regime exists at relatively low (but still finite) frequencies 
Therefore, we can expand $\Pi_{m}(E,\omega)$, Eq.~(\ref{34}), and keep only the leading terms, 
\begin{eqnarray}
\left |\Pi_{m}\left( E,\omega \right) \right|^{2} \approx   
\delta_{m,0} \left( 1 + \omega^{2}  {\rm Re} \frac{\partial^{2} \Pi_{m}\left( E,\omega \right) }{\partial \omega^{2} }  \Big |_{\omega = 0}
 \right)
\label{35} \\
+ \omega^{2} \left | \frac{\partial \Pi_{m}\left( E,\omega \right) }{\partial \omega }  \Big |_{\omega = 0}\right |^{2}
\,. \nonumber
\end{eqnarray}
Note, due to unitarity of the scattering matrix, Eq.~(\ref{uni}), the linear in $\omega$ term vanishes in  above equation.
As it follows quite generally from Eqs.~(\ref{33}) and (\ref{35}), the low frequency noise is quadratic in $\omega$, that is quite expected for a capacitor.  \cite{Buttiker:1993wh}

In the adiabatic regime we  use $ S_{F} \left( E_{q},E \right) \approx S_{q}(E)$ and $S_{F}^* \left( E_{q}+ \hbar \omega ,E_{m} + \hbar \omega \right) \approx S_{q-m}^{*}(E +  \hbar \omega)$ in equation (\ref{34}).
With these simplifications and expansion (\ref{35}) we find from Eq.~(\ref{33}) the following,
\begin{eqnarray}
{\cal P}_{0}^{(ex)}(\omega) &=& \frac{{\rm e}^{2} \hbar\omega^{2} }{2\pi} \int\limits_0^\infty  dE  \sum\limits_{m =  - \infty }^\infty  \left |\left ( S \frac{\partial S^{*} }{\partial E } \right )_{m}  \right |^{2} 
\label{36} \\ 
 &&\times \left\{    F ( {E ,E_{m}  + \hbar \omega } )   -   F ( {E ,E  + \hbar \omega } )   \right\} .
 \nonumber
\end{eqnarray} 
If the temperature is of the order of the frequency, $k_{B} T \sim \hbar\omega \sim \hbar\Omega$, then in the adiabatic regime the scattering amplitude can be kept constant over the entire energy interval relevant for the integration in above equation. 
That allow us to integrate over energy. 
With the scattering amplitude given in Eq.~(\ref{06-1}) we calculate, 
\begin{eqnarray}
{\cal P}_{0}^{(ex)}(\omega) = \frac{2 {\rm e}^{2} }{\pi } \left( \omega \tau_{D} \right)^{2} \left( \Omega \Gamma_{\tau} \right)^{2}   \sum\limits_{m =  - \infty }^\infty  e^{-2|m|\Omega\Gamma_{\tau}}  
\quad \quad 
\label{39} \\ 
\times \left\{  \left( m\Omega + \omega \right)  \coth\left( \frac{m\Omega + \omega }{2k_{B} T/\hbar } \right)   - \omega   \coth\left( \frac{\omega }{2k_{B} T/\hbar } \right)   \right\} ,
 \nonumber
\end{eqnarray} 
where the dwell time $\tau_{D} = h/({\rm T} \Delta )$.
Note, for the particular case, when the Fermi level aligns  with a quantum level in the cavity at $U=0$ and the time-dependent potential is ${\rm e}U(t) = (\Delta/2) \cos\left( \Omega t \right)$, we find the parameter $\Omega\Gamma_{\tau} = {\rm T}/(2\pi)$. 

\begin{figure}[b]
\begin{center}
\includegraphics[width=80mm]{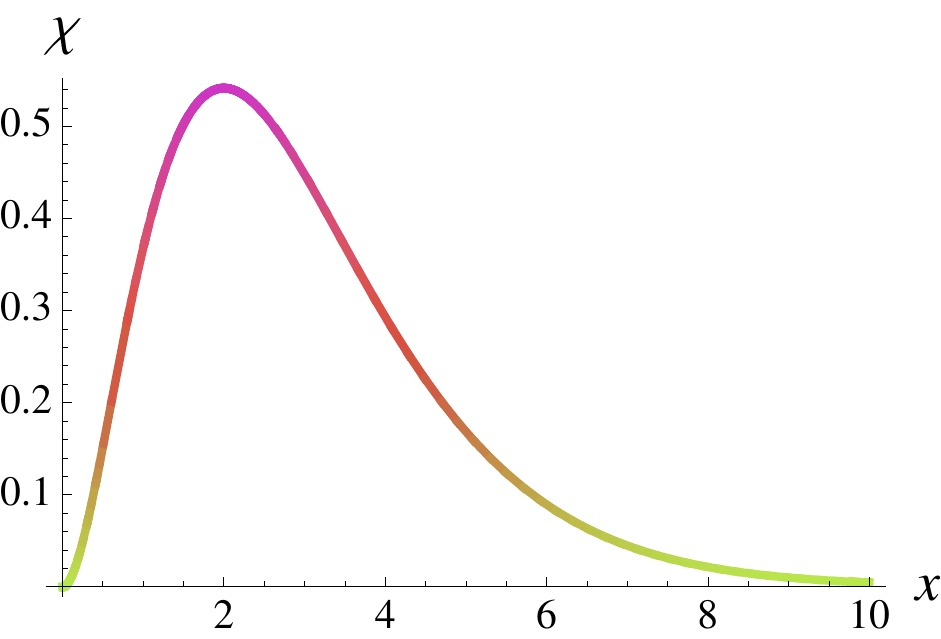}
\caption{(Color online)
Excess noise in the adiabatic regime as a function of frequency, ${\cal P}^{ex,ad}(\omega) = C \chi(2\omega\Gamma_{\tau})$, Eq.~(\ref{42}), with $C = 2({\rm e}^{2}/{\cal T }) \left[ \tau_{D} /(2\Gamma_{\tau} ) \right]^{2}$ and $\chi(x) = x^{2} e^{-x}$. The parameter $\Gamma_{\tau}$ is the half-width of an emitted current pulse. $\tau_{D}$ is the dwell time.}
\label{fig3}
\end{center}
\end{figure}

At zero temperature we sum up over $m$ in Eq.~(\ref{39}) and find,
\begin{eqnarray}
{\cal P}_{0}^{(ex)}(\omega) &=& 2\frac{{\rm e}^{2} }{\cal T } \left( \omega\tau_{D} \right)^{2} e^{- 2 |\omega| \Gamma_{\tau}} \,.
\label{42} 
\end{eqnarray}
This dependence is shown in Fig.~\ref{fig3}. 
At small frequencies, $\omega\Gamma_{\tau} \ll 1$, the noise power is quadratic in $\omega$. 
Such a quadratic dependence, found also for the cavity driven by a pulsed potential, is attributed to the probabilistic nature of tunneling of an electron leaving the cavity. \cite{Mahe:2010cp,Albert:2010co,Parmentier:2012ed}  
The factor $2$ in front of Eq.~(\ref{42}) refers to two particles, one electron and one hole, emitted during one period. 

At higher frequencies, $\omega\Gamma_{\tau} \gg 1$, the excess noise gets suppressed. 
As we already mentioned, this suppression is a consequence of boundedness  of a current spectrum, see Eq.(\ref{it04}). 
A complementary, energy-based interpretation of the excess noise suppression is also possible. 
It relies on a supposition \cite{note1} that an emitted electron does contribute to an excess noise power at frequency $\omega$ if it is able to emit an energy quantum $\hbar\omega$. 
 Taking into account that the energy of particles emitted by a single-electron source fluctuates, \cite{Battista:2012vs} we relate the factor $\exp(-2\omega\Gamma_{\tau})$ in Eq.~(\ref{42})  to the  number of electrons emitted with energy larger than $\hbar\omega$. 
We denote this number as ${\cal N}(\omega)$.  
Provided the energy probability distribution $p(E)$ for emitted electrons is known, this number can be calculated as follows,
\begin{eqnarray}
{\cal N} (\omega) = \int\limits_{\mu + \hbar\omega}^{\infty} dE p(E) \,.
\label{np}
\end{eqnarray}
To calculate $p(E)$ we note that in the quantized emission regime only a single particle can be emitted at a time. 
Therefore, the probability $p(E)$ can be calculated using the distribution function $f(E) = \left\langle \hat b^{\dag}(E) \hat b(E) \right\rangle$ for electrons scattered off the cavity. 
The difference between $p(E)$ and $f(E)$ is twofold. 
First, they are differently normalized. 
And, second, for $E < \mu$ the hole distribution function, $f^{h} = 1 - f$ has to be considered. 
Thus for $E>\mu$ we have, 
\begin{eqnarray}
p(E) = \frac{f(E) }{ \int\limits_{0}^{\infty} dE f (E) } \,.
\label{pf}
\end{eqnarray} 
To calculate $f(E)$ we use Eq.~(\ref{02add}) and find for the adiabatic regime and at zero temperature, 
\begin{eqnarray}
f(E) =  \sum_{q= \left [ (E - \mu)/(\hbar\Omega) \right]}^{\infty} \left|S_{q}\left(E\right) \right|^{2} \,.
\label{d04}
\end{eqnarray}
Here $\left [ \dots \right]$ stands for the integer part.
For the Breit-Wigner scattering amplitude, Eq.~(\ref{06-1}), the Fourier coefficients are, 
$S_{q} = -2\Omega \Gamma_{\tau} e^{- |q| \Omega \Gamma_{\tau} } e^{i q \Omega t_{\mp}} \theta(\pm q)$. 
Using these coefficients  we get $f(E)  =  2 \Omega \Gamma_{\tau} e^{-2\Omega\Gamma_{\tau}(n+1) }$ for $n\hbar\Omega < E - \mu < (n+1)\hbar\Omega$ with integer $n$. 
And, correspondingly, we calculate the energy probability distribution, 
\begin{eqnarray}
p (E) = \frac{2 \Gamma_{\tau}  }{ \hbar }  \, e^{ - \frac{2\Gamma_{\tau} }{\hbar }\, |E - \mu| }\,,
\label{pad}
\end{eqnarray}
and the average number of electrons contributing to the excess noise,
\begin{eqnarray}
{\cal N} (\omega) = e^{ - 2\Gamma_{\tau} |\omega | } \,.
\label{Nad}
\end{eqnarray}
This factor is exactly what enters Eq.~(\ref{42}).

\section{Conclusion}
\label{concl}

We presented here the scattering matrix approach to the non-stationary transport in coherent nanostructures. 
The essential feature of a non-stationary transport is an energy exchange between the carriers and the external driving force, that is naturally taken into account by the Floquet scattering matrix amplitudes dependent on two energies. 
Another advantage of the scattering formalism is that the results are presented in the form that admits an intuitive and transparent interpretation. 
For instance, the collective dynamics of electrons in a waveguide coupled to a driven cavity is naturally described in terms of extra particles, electrons and holes, emitted by the cavity.   

We use the scattering formalism to analyze the intrinsic noise, the phase noise, of a single-particle emitter recently investigated experimentally. 
The existence of this noise is rooted in the fundamental quantum-mechanical principle, the Heisenberg uncertainty principle. 
In the adiabatic emission regime analyzed here this relation manifests itself the most bright and the entire phase noise spectrum can be easily understood.   
The quadratic low frequency asymptotics is governed by the tunneling dynamics of an electron leaving the emitter: Before emission an electron occupies a quantum level in the cavity. Due to coupling between the cavity and an electron waveguide the quantum level acquires some width. Therefore, the initial energy of an electron has some uncertainty $2\delta$, which ultimately results in the uncertainty of the emission time, $\tau_{D} \sim \hbar/(2\delta)$. 
The latter in turn results in a quadratic phase noise spectrum at $\omega\tau_{D} \ll 1$. \cite{Mahe:2010cp,Albert:2010co,Parmentier:2012ed}
With increasing frequency another energy, the energy absorbed by an emitted electron (hole) from the external driving force \cite{Moskalets:2009dk}, comes into play. 
This energy ${\cal E}$ defines a time-extend, $2\Gamma_{\tau} = \hbar/{\cal E}$, of an emitted single-particle state. \cite{Olkhovskaya:2008en,Haack:2011em} 
The finite duration of a current pulse bounds the frequency spectrum of a current and, therefore, the one of the phase noise.

\begin{acknowledgements}
I thank Mathias Albert
and Markus B\"{u}ttiker for helpful discussions and valuable comments on the manuscript. 
I thank the University of Geneva for warm hospitality, where part of this work was carried out. 
\end{acknowledgements}


\begin{thebibliography}{}
%
%

\bibitem{Moskalets:2011cw}
Moskalets, M.V.: Scattering matrix approach to non-stationary quantum transport. Imperial College Press, London (2011). 

\bibitem{Blumenthal07}
Blumenthal, M. D.,  Kaestner, B.,  Li, L.,  Giblin, S.,  Janssen, T. J. B. M.,  Pepper, M.,  Anderson,  D.,  Jones, G., \&  Ritchie, D. A.: Gigahertz quantized charge pumping. 
Nature Physics. {\bf 3}, 343--347 (2007).

\bibitem{Feve:2007jx}
F\`{e}ve, G.,  Mah\'e, A.,  Berroir, J.-M.,  Kontos, T.,  Pla\c{c}ais, B.,  Glattli, D. C.,  Cavanna, A.,  Etienne, B.,  Jin, Y.: An on-demand coherent single-electron source.  
Science. {\bf 316}, 1169--1172 (2007).

\bibitem{Kaestner:2008gv}
Kaestner, B.,  Kashcheyevs, V.,  Amakawa, S.,  Blumenthal, M.,  Li, L.,  Janssen, T. J. B. M.,  Hein, G.,  Pierz, K.,  Weimann, T.,  Siegner, U., \& Schumacher, H. W.: Single-parameter nonadiabatic quantized charge pumping. 
Physical Review B. {\bf 77}, 153301 (2008).

\bibitem{Fujiwara:2008gt}
Fujiwara, A.,  Nishiguchi, K., \&  Ono, Y.: Nanoampere charge pump by single-electron ratchet using silicon nanowire metal-oxide-semiconductor field-effect transistor. 
Appl. Phys. Lett. {\bf 92}, 042102 (2008).

\bibitem{Dubois:2013ul}
Dubois, J., Jullien, T., Roulleau, P., Portier, F., Roche, P., Cavanna, A., Jin, Y., Wegscheider, W., Glattli, D.C., (in preparation).

\bibitem{Giblin:2012cl}
Giblin, S.P., Kataoka, M., Fletcher, J.D., See, P., Janssen, T.J.B.M., Griffiths, J.P., Jones, G.A.C., Farrer, I., \& Ritchie, D.A.: Towards a quantum representation of the ampere using single electron pumps. Nature Communications. {\bf 3}, 930 (2012).

\bibitem{Hohls:2012gf}
Hohls, F., Welker, A.C., Leicht, C., Fricke, L., Kaestner, B., Mirovsky, P., M\"uller, A., Pierz, K., Siegner, U., \& Schumacher, H.W.: Semiconductor Quantized Voltage Source. Physical Review Letters. {\bf 109}, 56802 (2012).

\bibitem{Pekola:2012ti}
Pekola, J.P. \& Saira, O.P., Maisi, V.F., Kemppinen, A., M\"ott\"onen, M., Pashkin, Y.A., Averin, D.V.: Single-electron current sources: towards a refined definition of ampere. arXiv:1208.4030v1 (unpublished).

\bibitem{Jehl:2013wd}
Jehl, X., Voisin, B., Charron, T., Clapera, P., Ray, S., Roche, B., Sanquer, M., Djordjevic, S., Devoille, L., Wacquez, R., \& Vinet, M.: A hybrid metal/semiconductor electron pump for practical realization of a quantum ampere, arXiv:1302.6470v1 (unpublished).

\bibitem{Likharev:1985ul}
Likharev, K. \& Zorin, A.B.: Theory of the Bloch-wave oscillations in small Josephson junctions. Journal of low temperature physics. {\bf 59}, 347--382 (1985).

\bibitem{Flowers:2004ei}
Flowers, J.: The Route to Atomic and Quantum Standards. Science. {\bf 306}, 1324--1330 (2004).

\bibitem{Feltin:2009fo}
Feltin, N. \& Piquemal, F.: Determination of the elementary charge and the quantum metrological triangle experiment. Eur. Phys. J. Spec. Top. {\bf 172}, 267--296 (2009).

\bibitem{Scherer:2012fe}
Scherer, H. \& Camarota, B.: Quantum metrology triangle experiments: a status review. Meas. Sci. Technol. {\bf 23}, 124010 (2012).

\bibitem{Grenier:2011js}
Grenier, C. \& Herv\'e, R., F\`{e}ve, G., Degiovanni, P.: Electron quantum optics in quantum Hall edge channels. Int. J. Mod. Phys. B. {\bf 25}, 1053--1073 (2011).

\bibitem{Bennett:2000kl}
Bennett, C.H., DiVincenzo, D.P.: Quantum information and computation. Nature. {\bf 404}, 247--255 (2000).

\bibitem{Olkhovskaya:2008en}
Ol'khovskaya, S., Splettstoesser, J., Moskalets, M., \& B\"{u}ttiker, M.: Shot Noise of a Mesoscopic Two-Particle Collider. Physical Review Letters. {\bf 101}, 166802 (2008).

\bibitem{Bocquillon:2013dp}
Bocquillon, E., Freulon, V., Berroir, J.-M., Degiovanni, P.,  Pla\c{c}ais, B., Cavanna, A., Jin, Y., F\`{e}ve, G.: Coherence and Indistinguishability of Single Electrons Emitted by Independent Sources. Science. {\bf 339}, 1054--1057 (2013).

\bibitem{Mahe:2010cp}
Mah\'e, A., Parmentier, F.D., Bocquillon, E., Berroir, J.-M., Glattli, D., Kontos, T., Pla\c{c}ais, B., F\`{e}ve, G., Cavanna, A., \& Jin, Y.: Current correlations of an on-demand single-electron emitter. Physical Review B. {\bf 82}, 201309(R) (2010).

\bibitem{Albert:2010co}
Albert, M., Flindt, C., \& B\"{u}ttiker, M.: Accuracy of the quantum capacitor as a single-electron source. Physical Review B. {\bf 82}, 041407(R) (2010).

\bibitem{Parmentier:2012ed}
Parmentier, F.D., Bocquillon, E., Berroir, J.-M., Glattli, D., Pla\c{c}ais, B., \& F\`{e}ve, G., Albert, M., Flindt, C., \& B\"{u}ttiker, M.: Current noise spectrum of a single-particle emitter: Theory and experiment. Physical Review B. {\bf 85}, 165438 (2012).

\bibitem{Moskalets:2002hu}
Moskalets, M.  \& B\"{u}ttiker, M.: Floquet scattering theory of quantum pumps. Physical Review B. {\bf 66}, 205320 (2002). 

\bibitem{Landauer:1988vs}
Landauer, R.: Spatial variation of currents and fields due to localized scatterers in metallic conduction. IBM J. Res. \& Dev. {\bf 32}, 306--316 (1988).

\bibitem{Switkes:1999fj}
Switkes, M., Marcus, C.M., Campman, K., Gossard, A.C.: An Adiabatic Quantum Electron Pump. Science. {\bf 283}, 1905--1908 (1999).

\bibitem{Platero:2004ep}
Platero, G., Aguado, R.: Photon-assisted transport in semiconductor nanostructures. Physics Reports. {\bf 395}, 1--157 (2004).

\bibitem{Arrachea:2007ew}
Arrachea, L.: Exact Green's function renormalization approach to spectral properties of open quantum systems driven by harmonically time-dependent fields. Physical Review B. {\bf 75}, 035319 (2007).

\bibitem{Moskalets:2004bo}
Moskalets, M. \& B\"{u}ttiker, M.: Adiabatic quantum pump in the presence of external ac voltages. Physical Review B. {\bf 69}, 205316 (2004).

\bibitem{Buttiker:1992vr}
B\"{u}ttiker, M.: Scattering theory of current and intensity noise correlations in conductors and wave guides. Physical Review B. {\bf 46}, 12485--12507 (1992).

\bibitem{Blanter:2000wi}
Blanter, Y.M., B\"{u}ttiker, M.: Shot noise in mesoscopic conductors. Physics Reports. {\bf 336}, 1--166 (2000).

\bibitem{Buttiker:1993wh}
B\"{u}ttiker, M., Thomas, H., \& Pr\^{e}tre, A.: Mesoscopic capacitors. Physics Letters A. {\bf 180}, 364--369 (1993).

\bibitem{Klitzing:1980wf}
Klitzing, K.  \& Dorda, G.  \& Pepper, M.: New method for high-accuracy determination of the fine-structure constant based on quantized Hall resistance. Physical Review Letters. {\bf 45}, 494--497 (1980).

\bibitem{Halperin:1982tb}
Halperin, B.I.: Quantized Hall conductance, current-carrying edge states, and the existence of extended states in a two-dimensional disordered potential. Physical Review B. {\bf 25}, 2185 (1982).

\bibitem{Buttiker:1988vk}
B\"{u}ttiker, M.: Absence of backscattering in the quantum Hall effect in multiprobe conductors. Physical Review B. {\bf 38}, 9375 (1988).

\bibitem{Pretre:1996uw}
Pr\^{e}tre, A., Thomas, H., \& B\"{u}ttiker, M.: Dynamic admittance of mesoscopic conductors: Discrete-potential model. Physical Review B. {\bf 54}, 8130 (1996).

\bibitem{Gabelli:2006eg}
Gabelli, J., F\`{e}ve, G., Berroir, J.-M., Pla\c{c}ais, B., Cavanna, A., Etienne, B., Jin, Y., Glattli, D.: Violation of Kirchhoff's Laws for a Coherent RC Circuit. Science. {\bf 313}, 499--502 (2006).

\bibitem{Moskalets:2008fz}
Moskalets, M., Samuelsson, P. \& B\"uttiker, M. Quantized Dynamics of a Coherent Capacitor. Physical Review Letters.  {\bf 100}, 086601 (2008). 

\bibitem{Moskalets:2008ii}
Moskalets, M. \& B\"uttiker, M.: Dynamic scattering channels of a double barrier structure. Physical Review B. {\bf 78}, 035301 (12) (2008).

\bibitem{Moskalets:2007dl}
Moskalets, M. \& B\"{u}ttiker, M.: Time-resolved noise of adiabatic quantum pumps. Physical Review B. {\bf 75}, 035315 (2007).

\bibitem{Gardiner:2000wq}
Gardiner, C.W., Zoller, P.: Quantum Noise. Springer, New York (2000).

\bibitem{Clerk:2010dh}
Clerk, A.A., Girvin, S.M., Marquardt, F., Schoelkopf, R.J.: Introduction to quantum noise, measurement, and amplification. Rev. Mod. Phys. {\bf 82}, 1155--1208 (2010).

\bibitem{Breit:1936ud}
Breit, G.  \& Wigner, E.: Capture of slow neutrons. Phys. Rev. {\bf 49}, 519 (1936).


\bibitem{Splettstoesser:2008gc}
Splettstoesser, J., Ol'khovskaya, S., Moskalets, M., \& B\"{u}ttiker, M.: Electron counting with a two-particle emitter. Physical Review B. {\bf 78}, 205110 (2008).

\bibitem{note1}
This supposition seems to be applicable also to the case of a DC biased quantum point contact with non-perfect transmission \cite{Blanter:2000wi,Aguado:2000vm}. 

\bibitem{Aguado:2000vm}
Aguado, R. \& Kouwenhoven, L.P.: Double quantum dots as detectors of high-frequency quantum noise in mesoscopic conductors. Physical Review Letters. {\bf 84}, 1986--1989 (2000).


\bibitem{Battista:2012vs}
Battista, F., Moskalets, M., Albert, M., \& Samuelsson, P.: Quantum heat fluctuations of single particle sources. Physical Review Letters. {\bf 110}, 126602 (2013).

\bibitem{Moskalets:2009dk}
Moskalets, M. \& B\"{u}ttiker, M.: Heat production and current noise for single- and double-cavity quantum capacitors. Physical Review B. {\bf 80}, 081302(R) (2009).

\bibitem{Haack:2011em}
Haack, G., Moskalets, M., Splettstoesser, J., \& B\"{u}ttiker, M.: Coherence of single-electron sources from Mach-Zehnder interferometry. Physical Review B. {\bf 84}, 081303 (2011).




\end{thebibliography}


\end{document}